\PassOptionsToPackage{unicode}{hyperref}
\PassOptionsToPackage{hyphens}{url}
\documentclass[
  10pt,
]{article}
\usepackage{amsmath,amssymb}
\usepackage{iftex}
\ifPDFTeX
  \usepackage[T1]{fontenc}
  \usepackage[utf8]{inputenc}
  \usepackage{textcomp} 
\else 
  \usepackage{unicode-math} 
  \defaultfontfeatures{Scale=MatchLowercase}
  \defaultfontfeatures[\rmfamily]{Ligatures=TeX,Scale=1}
\fi
\usepackage{lmodern}
\ifPDFTeX\else
\fi
\IfFileExists{upquote.sty}{\usepackage{upquote}}{}
\IfFileExists{microtype.sty}{
  \usepackage[]{microtype}
  \UseMicrotypeSet[protrusion]{basicmath} 
}{}
\makeatletter
\@ifundefined{KOMAClassName}{
  \IfFileExists{parskip.sty}{%
    \usepackage{parskip}
  }{
    \setlength{\parindent}{0pt}
    \setlength{\parskip}{6pt plus 2pt minus 1pt}}
}{
  \KOMAoptions{parskip=half}}
\makeatother
\usepackage{xcolor}
\usepackage[margin=0.85in]{geometry}
\usepackage{longtable,booktabs,array}
\usepackage{calc} 
\usepackage{etoolbox}
\makeatletter
\patchcmd\longtable{\par}{\if@noskipsec\mbox{}\fi\par}{}{}
\makeatother
\IfFileExists{footnotehyper.sty}{\usepackage{footnotehyper}}{\usepackage{footnote}}
\makesavenoteenv{longtable}
\usepackage{graphicx}
\makeatletter
\def\maxwidth{\ifdim\Gin@nat@width>\linewidth\linewidth\else\Gin@nat@width\fi}
\def\maxheight{\ifdim\Gin@nat@height>\textheight\textheight\else\Gin@nat@height\fi}
\makeatother
\setkeys{Gin}{width=\maxwidth,height=\maxheight,keepaspectratio}
\makeatletter
\def\fps@figure{htbp}
\makeatother
\ifLuaTeX
  \usepackage{selnolig}  
\fi
\usepackage[]{biblatex}
\usepackage{bookmark}
\IfFileExists{xurl.sty}{\usepackage{xurl}}{} 
\hypersetup{
  pdftitle={Why Three Phases? A Historical and Engineering Reassessment of Phase Order in AC Power Transmission},
  pdfauthor={Kai Sun},
  hidelinks,
  pdfcreator={LaTeX via pandoc}}

\title{Why Three Phases? A Historical and Engineering Reassessment of
Phase Order in AC Power Transmission}
\author{Kai Sun\\
\small Department of Electrical Engineering and Computer Science\\
\small University of Tennessee, Knoxville, TN, USA\\
\small \texttt{kaisun@utk.edu}
}
\date{}

\begin{document}
\maketitle
\begin{abstract}
Three-phase alternating current (AC) is so deeply embedded in modern
electric-power infrastructure that its phase order is often treated as
self-evident. Historically, however, single-phase and true two-phase
systems were commercially important, while six-phase high-phase-order
transmission was later demonstrated on an operating utility system. This
paper reassesses why three phases became the dominant architecture for
bulk AC transmission. A general balanced \(m\)-phase formulation is used
to show that constant aggregate instantaneous power is not unique to
three phases: an ideal balanced two-phase system also cancels the
double-frequency power term and can generate a constant-magnitude
rotating field. Consequently, the historical displacement of two phases
cannot be explained by power smoothness alone. The comparison is instead
organized around conductor architecture, insulation stress, machine and
transformer utilization, conversion requirements, right-of-way
utilization, and technological path dependence. A historical, commercial demonstration of the NYSEG Goudey--Oakdale six-phase project has shown that
high-phase-order transmission was technically feasible and could improve
corridor utilization. The paper then asks a forward-looking question: if
power-electronic conversion and protection make phase count less costly,
could \(m>3\) offer intrinsic advantages? Six such advantages are
identified: modular decomposition into interleaved three-phase groups,
redundant control degrees of freedom, structured modal/fault analysis,
increased corridor power density under field constraints, potentially
higher natural loading and loadability, and enhanced harmonic/field
cancellation. A companion derivation shows that phase count alone does
not intrinsically reduce \(I^2R\) loss at fixed total conductor material
and phase voltage. The resulting conclusion is therefore conditional:
three phases are an unusually favorable historical optimum for the
electromechanical grid, but not a mathematically universal optimum for a
future converter-dominated grid.
\end{abstract}

\section{I. Introduction}\label{i.-introduction}

The number of phases in an AC power system is an architectural choice
that reaches far beyond notation. Phase order determines the angular
arrangement of voltages and currents, the structure of rotating magnetic
fields, the number and size of line conductors, transformer and machine
windings, switching poles, protection measurements, fault states, and
the dimensionality of network and control models. Yet modern power
engineering usually begins after this choice has already been made:
generators are assumed to be three-phase, transmission circuits have
three-phase conductors, and the familiar relation
\(P_\Sigma=\sqrt{3}V_{LL}I_L\cos\phi\) is introduced as a basic identity rather
than as the result of one historical design choice among several.

That conventional starting point hides an instructive engineering
question. In the late nineteenth century, practical single-phase and
two-phase systems existed alongside emerging three-phase systems.
Niagara Falls initially used large two-phase generators, and
Scott-connected transformers converted their output to three phases for
the transmission link to Buffalo \autocite{ethwBuffalo,ethwScott}. More
than a century later, high-phase-order (HPO) research showed that
six-phase transmission could be integrated into an operating three-phase
utility network \autocite{brown1991,nyseg1997}. Three phases therefore
did not become universal because other phase orders were physically
impossible.

The literature addressing this question is fragmented. Histories of
two-phase technology emphasize its commercial importance and eventual
displacement \autocite{blalock2004}. Historical studies of early
three-phase systems examine the rapid development of three-phase
generation, transformation, transmission, and motors in the 1890s
\autocite{allerhand2020}. A separate body of research analyzes six- and
twelve-phase HPO transmission for corridor compaction, insulation,
economics, and protection
\autocite{stewart1978a,hpoecon1984,landers1998}. Still another
literature treats multiphase machines and converters, where more than
three phases may be desirable for fault tolerance, power density, or
current sharing \autocite{levi2008}. These strands are rarely organized
around one common phase-order question.

This paper provides a unified analytical, historical, and engineering
synthesis. It does not claim novelty for standard polyphase identities.
Instead, it uses a general \(m\)-phase framework to clarify which
properties are fundamental consequences of phase order and which are
consequences of implementation choices. Six points are emphasized.
First, constant aggregate instantaneous power is not unique to three
phases. Second, conductor-economy claims depend on what voltage,
insulation stress, power, and loss constraints are held fixed. Third,
three phases have a particularly economical symmetric three-conductor
realization for balanced bulk power. Fourth, six-phase HPO transmission
is a real demonstrated technology whose benefits are principally
corridor-specific rather than a universal replacement for three phases.
Fifth, eliminating selected occurrences of \(\sqrt{3}\) does not by
itself make an HPO system simpler. Sixth, that last
statement must be qualified for a future converter-dominated grid: additional phases can create useful structure and controllable redundancy, so the
optimal phase order can change when phase conversion, switching,
sensing, and fault accommodation are implemented predominantly by power
electronics.

\section{II. Historical Competition and Engineering Comparison Among Phase Orders}\label{ii.-historical-competition-among-phase-orders}

\subsection{A. Single-phase transmission}\label{a.-single-phase-transmission-was-practical}

Early AC transmission established that voltage transformation could
reduce current and therefore conductor loss over distance. An IEEE
Milestone account of the Ames hydroelectric installation near Telluride,
Colorado, records a 100-hp Westinghouse single-phase alternator
generating at approximately 3000 V and 133 Hz and supplying a motor
about 2.6 miles away \autocite{ethwAmes}. The system was not a
laboratory curiosity: it transmitted useful mechanical power to a mining
operation. Portland General Electric's historical study records that 4,000-V
Westinghouse single-phase AC equipment at Willamette Falls was transmitting
power commercially to Portland by late 1890, making it another important
early long-distance AC installation \autocite{pge2008}.

Consider sinusoidal voltages and currents,
\begin{equation}
v(t)=\sqrt{2}V\cos\omega t, \qquad i(t)=\sqrt{2}I\cos(\omega t-\phi),
\label{eq:single-phase-vi}
\end{equation}
where \(V\) and \(I\) are RMS quantities defined as the respective peak voltage and current divided by \(\sqrt2\), and \(\phi\) is the power factor angle. Using \(2\cos A\cos B=\cos(A-B)+\cos(A+B)\), the instantaneous power is
\begin{equation}
\begin{aligned}
p(t)&=v(t)i(t) \\
&=VI\cos\phi+VI\cos(2\omega t-\phi).
\end{aligned}
\label{eq:single-phase-power}
\end{equation}
The first term is average real power and the second oscillates at twice
electrical frequency. The machine or load must exchange the associated
energy cyclically with electrical, magnetic, mechanical, or storage
elements. This is not fatal since single-phase systems remain
indispensable, but it is an unnecessary complication for large rotating
generation and motor systems when a balanced polyphase supply is
available \autocite{chapman2012,fitzgerald2003}.

Single-phase AC transmission therefore deserves to be regarded as an engineering
baseline rather than as a failed precursor. Its enduring advantages are
obvious: two line conductors suffice, transformers are straightforward,
and a vast range of lighting and small loads can be served efficiently.
Its weakness for bulk electromechanical power is not inability to
transmit energy, but its pulsating instantaneous power and its inability to naturally produce a self-starting rotating magnetic field with a single stator phase. An analogy is trying to maintain a bicycle at constant speed by pedaling with only one foot.

\subsection{B. Two-phase transmission}\label{b.-two-phase-was-a-serious-commercial-architecture}

True two-phase power consists of two sinusoidal phase sets in time
quadrature, conventionally separated by \(90^\circ\). This should not be
confused with North American 120/240-V split-phase service, whose two
hot conductors are opposite ends of one single-phase secondary winding
and are \(180^\circ\) apart relative to the center tap. Assume

\begin{equation}
v_\alpha(t)=\sqrt{2}V\cos\omega t,
\qquad
i_\alpha(t)=\sqrt{2}I\cos(\omega t-\phi),
\label{eq:two-phase-alpha}
\end{equation}
and
\begin{equation}
v_\beta(t)=\sqrt{2}V\sin\omega t,
\qquad
i_\beta(t)=\sqrt{2}I\sin(\omega t-\phi).
\label{eq:two-phase-beta}
\end{equation}

The phase powers are
\begin{equation}
p_\alpha(t)=VI\left[\cos\phi+\cos(2\omega t-\phi)\right],
\qquad
p_\beta(t)=VI\left[\cos\phi-\cos(2\omega t-\phi)\right].
\label{eq:two-phase-powers}
\end{equation}
Hence
\begin{equation}
p_\Sigma(t)=2VI\cos\phi.
\label{eq:two-phase-total-power}
\end{equation}

The total is constant and is exactly equal to the total real power being delivered by both phases. Likewise, two spatial windings displaced by
\(90^\circ\) and excited by quadrature currents can create a
constant-magnitude rotating magnetic field. The common assertion that
two-phase power lost because its balanced total power pulsates is
therefore incorrect. The historical competition must be explained
elsewhere.

A conventional true two-phase distribution system can be implemented
with four wires, two for each independent phase circuit. A three-wire
variant shares a common conductor. If the two phase currents have equal
RMS magnitude \(I\) and are \(90^\circ\) apart, the common-conductor
current is

\begin{equation}
I_c=|\mathbf I_\alpha+\mathbf I_\beta|=\sqrt{2}I,
\label{eq:two-phase-common-current}
\end{equation}

so conductor sharing does not create the zero-current neutral familiar
from a balanced three-phase system.

Two-phase systems were commercially important. Westinghouse equipment
for the 1893 World's Columbian Exposition and the Niagara Falls project
belonged to this period of genuine competition among polyphase
architectures \autocite{blalock2004}. At Niagara, the initial generating
units were two-phase machines. IEEE historical documentation describes
2200-V, 25-Hz two-phase generation and local two-phase distribution
\autocite{ethwBuffalo}. This architecture could run polyphase motors and
could supply single-phase loads conveniently from either phase pair.

The most revealing feature of the Niagara Falls project is that the same project did not
use one phase order for every function. For transmission to Buffalo, two
single-phase transformer units were Scott connected to convert 2200-V
two-phase power to an 11-kV three-phase, three-wire line. In \autocite{ethwBuffalo}, Charles F. Scott explicitly described his
phase-transformation method (so called "Scott connection") as a way to combine advantages then
associated with two-phase distribution and three-phase transmission
\autocite{ethwScott}. Service to Buffalo began in 1896. Thus one of the
canonical two-phase generating stations in power-system history
simultaneously provides early evidence of the attraction of three-wire
three-phase transmission.

\subsection{C. Three-phase transmission}\label{c.-three-phase-emerged-rapidly}

A balanced positive-sequence three-phase set can be written

\begin{equation}
v_a(t)=\sqrt{2}V\cos\omega t, \qquad
v_b(t)=\sqrt{2}V\cos(\omega t-2\pi/3),\qquad
v_c(t)=\sqrt{2}V\cos(\omega t+2\pi/3).
\label{eq:three-phase-voltages}
\end{equation}

Similar to two-phase transformation, it can easily be proved that the double-frequency components also sum to zero, yielding

\begin{equation}
p_\Sigma(t)=3VI\cos\phi.
\label{eq:three-phase-total-power}
\end{equation}
which is the total real power delivered by three phases, namely,

\begin{equation}
P_\Sigma=3P_{ph}=3V_{ph}I_{ph}\cos\phi.
\label{eq:three-phase-power-phase}
\end{equation}

For a balanced wye system, a line-to-line voltage is the phasor
difference \(\mathbf V_{ab}=\mathbf V_a-\mathbf V_b\). Because the two
phase-to-neutral phasors are separated by \(120^\circ\),

\begin{equation}
\begin{aligned}
V_{LL}
&=V_{ph}\sqrt{1+1-2\cos120^\circ}\\
&=\sqrt{3}V_{ph}.
\end{aligned}
\label{eq:three-phase-vll}
\end{equation}

Thus, there is also
\begin{equation}
P_\Sigma=\sqrt{3}V_{LL}I_L\cos\phi.
\label{eq:three-phase-power-line}
\end{equation}

The factor \(\sqrt{3}\) is a chord length in phasor geometry, not an
inefficiency or computational burden in modern computation.

The Lauffen--Frankfurt demonstration of 1891 is a landmark in the
history of long-distance three-phase transmission, and Allerhand's
historical review places it within a remarkably rapid sequence of
three-phase developments between 1891 and 1893 \autocite{allerhand2020}.
These systems joined three-phase generation, transformation,
transmission, and utilization into an integrated architecture. The
importance of this integration should not be understated. A technically
superior line topology is of limited value if it creates expensive
interfaces with generators, transformers, and motors; three phases proved
to be attractive at the level of the entire electromechanical system. Over the following decades, manufacturers increasingly standardized machines, transformers, switchgear, protection, and operating practice around three phases. The phase
order became not only an electrical design variable but also an
industrial standard with strong network effects.

\subsection{D. Why Three Phases Defeated Two
Phases}\label{v.-why-three-phases-defeated-two-phases}

\begin{figure}
\centering
\includegraphics[width=0.88\textwidth,height=\textheight]{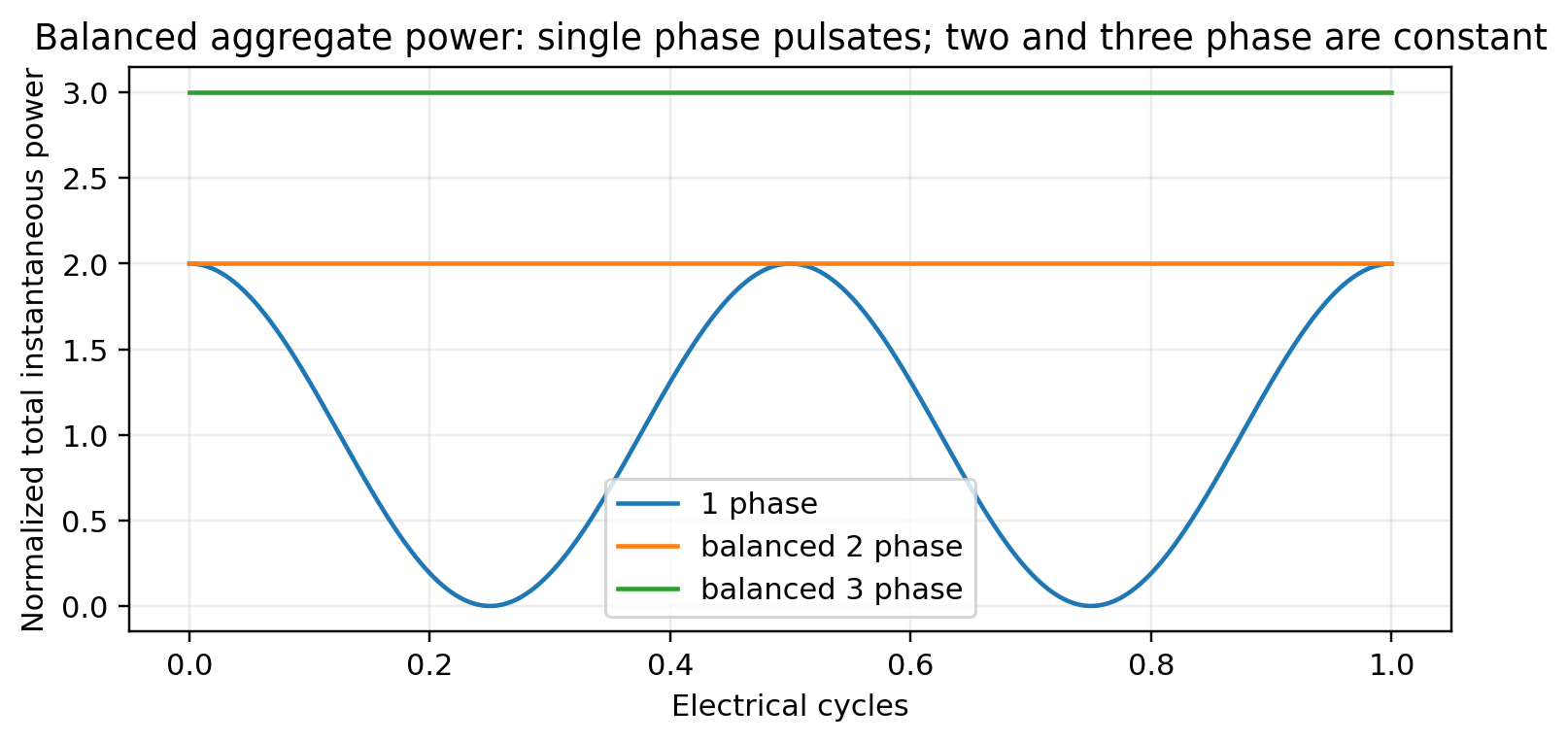}
\caption{Normalized aggregate instantaneous power. The single-phase
curve is shown at unity average power; balanced two- and three-phase
totals are constant under the stated assumptions.}\label{fig:power}
\end{figure}

Both ideal two-phase and three-phase systems can deliver constant
balanced power and create a rotating field as illustrated in Fig. \ref{fig:power}. Two- and three-phase machines can both generate rotating fields. The
advantage of three phase is therefore not existence of rotation but
integration and utilization. 

At the equipment level, three stator windings spaced \(120^\circ\)
apart provide cyclic symmetry and connect naturally in wye or delta.
Standard machine design has exploited this architecture for more than a
century \autocite{chapman2012,fitzgerald2003}. Three-phase transformer
banks can be assembled from single-phase units or integrated into
three-phase cores, while wye/delta connections provide useful voltage and grounding options, while a closed delta winding provides a path for triplen-harmonic currents. Actually, two-phase equipment can be equally elegant within its own ecosystem, and
the Scott connection \autocite{ethwScott} shows that two and three phase can be
interconnected without rotating conversion. But every phase-order
interface is additional apparatus. Once three-phase transmission and
three-phase motors grew rapidly, this interface burden became an
economic argument for convergence on one standard.

At the system level, the relevant distinction is economy, not power smoothness. A strong engineering distinction is that three phase obtains those properties in a symmetric 
three-wire bulk-power circuit. For balanced phase currents,
\begin{equation}
i_a+i_b+i_c=0,
\label{eq:three-phase-current-sum}
\end{equation}
so the bulk transmission path requires no neutral. Each line conductor
is equivalent under cyclic phase permutation. The symmetry propagates
naturally into generator windings, transformer connections, line
transposition, and machine stators. A conventional four-wire two-phase circuit, in contrast, allocates a
complete conductor pair to each orthogonal phase. A three-wire two-phase
circuit can share a common conductor, but \eqref{eq:two-phase-common-current} shows that the shared
conductor is not unloaded under balanced operation. The result is not
that a two-phase line cannot be engineered economically; rather, it
lacks the unusually clean three-conductor symmetry of three phase.

Textbooks and historical accounts often quote simple copper ratios for
single-, two-, and three-phase systems. Such numbers can be useful, but
only after the comparison basis is specified. The distinction in conductor economy between two-phase transmission and three-phase transmission can be inspected with explicit comparison assumptions. A fair transmission
comparison may hold fixed any combination of delivered real power,
conductor loss, maximum conductor-to-ground voltage, maximum
conductor-to-conductor voltage, conductor material, temperature rise, or
insulation coordination. Different constraints produce different
numerical ratios.

A deliberately simplified loss-based calculation illustrates the issue.
For \(n\) identical conductors of length \(\ell\), area \(A\), and
resistivity \(\rho\), total conductor loss is

\begin{equation}
P_{loss}=nI^2\frac{\rho\ell}{A}.
\label{eq:conductor-loss-general}
\end{equation}

If systems are compared at equal real power \(P_\Sigma\), equal power factor,
equal numerical working voltage in the respective power equations, and
equal total \(I^2R\) loss, a four-wire two-phase system has per-phase
current \(P_\Sigma/(2V\cos\phi)\), while a three-phase system expressed at
line-to-line voltage has current \(P_\Sigma/(\sqrt{3}V\cos\phi)\). Under that
particular definition, the familiar result that the three-phase
arrangement can require less total conductor cross-section follows. But
the comparison is not invariant to a change in voltage constraint.

For example, if the limiting insulation quantity is conductor-to-ground
voltage rather than the numerical line-to-line voltage used in the
preceding formulas, the permissible voltages of the competing systems
scale differently. If the limiting quantity is maximum phase-to-phase
stress, a two-phase \(90^\circ\) phasor difference and a three-phase
\(120^\circ\) difference again lead to different voltage utilization.
Real overhead-line optimization further depends on corona, electric
field, surge impedance, mechanical sag, bundle geometry, and contingency
criteria \autocite{grainger1994}. Therefore this paper deliberately
avoids presenting one conductor percentage as a universal theorem.

The robust conclusion is qualitative but strong: three phase combines
balanced constant power with a fully symmetric three-conductor
realization in which no dedicated or heavily loaded common return
conductor is required. That property is independent of the exact
copper-economy metric.

\begin{longtable}[]{@{}
  >{\raggedright\arraybackslash}p{(\columnwidth - 6\tabcolsep) * \real{0.2500}}
  >{\raggedright\arraybackslash}p{(\columnwidth - 6\tabcolsep) * \real{0.2500}}
  >{\raggedright\arraybackslash}p{(\columnwidth - 6\tabcolsep) * \real{0.2500}}
  >{\raggedright\arraybackslash}p{(\columnwidth - 6\tabcolsep) * \real{0.2500}}@{}}
\caption{Comparison of basic phase-order characteristics. Numerical
conductor-economy ratios require additional assumptions.}\tabularnewline
\toprule\noalign{}
\begin{minipage}[b]{\linewidth}\raggedright
Criterion
\end{minipage} & \begin{minipage}[b]{\linewidth}\raggedright
Single phase
\end{minipage} & \begin{minipage}[b]{\linewidth}\raggedright
True two phase
\end{minipage} & \begin{minipage}[b]{\linewidth}\raggedright
Three phase
\end{minipage} \\
\midrule\noalign{}
\endfirsthead
\toprule\noalign{}
\begin{minipage}[b]{\linewidth}\raggedright
Criterion
\end{minipage} & \begin{minipage}[b]{\linewidth}\raggedright
Single phase
\end{minipage} & \begin{minipage}[b]{\linewidth}\raggedright
True two phase
\end{minipage} & \begin{minipage}[b]{\linewidth}\raggedright
Three phase
\end{minipage} \\
\midrule\noalign{}
\endhead
\bottomrule\noalign{}
\endlastfoot
Typical balanced bulk-power conductors & 2 & 4 (or 3 with shared
conductor) & 3 \\
Constant aggregate power, balanced sinusoidal case & No & Yes & Yes \\
Natural rotating field with polyphase stator & Not from one phase alone
& Yes & Yes \\
Balanced neutral/common-conductor current & Not applicable & Shared
conductor generally nonzero in 3-wire form & Zero in ideal 3-wire set \\
Symmetric line-conductor roles & Yes, two-wire pair & Less direct in
common 3-wire form & Yes, cyclic symmetry \\
Mature bulk-grid equipment ecosystem & Specialized/limited &
Legacy/niche & Universal standard \\
\end{longtable}

\subsection{E. Six-phase transmission}\label{vii.-six-phase-high-phase-order-transmission-a-real-world-case-study}

Six-phase high phasor order (HPO) transmission was developed as a corridor-utilization concept rather
than as a proposal to rebuild all generators as six-phase machines.
Stewart and Wilson's 1978 feasibility work analyzed phase orders above
the conventional three-phase arrangement as a way to use overhead
transmission rights-of-way more efficiently \autocite{stewart1978a}.
Later studies addressed insulation, environmental performance,
economics, and practical conversion of existing double-circuit
structures \autocite{stewart1992,hpoecon1984}. A particularly attractive retrofit case is an existing double-circuit
three-phase corridor, which already contains six phase conductors.
Reconnecting and physically ordering those conductors as one symmetrical
six-phase circuit changes the electrical relationship among adjacent
conductors without necessarily increasing the conductor count. The line
can then be interfaced to the surrounding three-phase grid through
phase-conversion transformers.

\begin{figure}
\centering
\includegraphics[width=0.9\textwidth,height=\textheight]{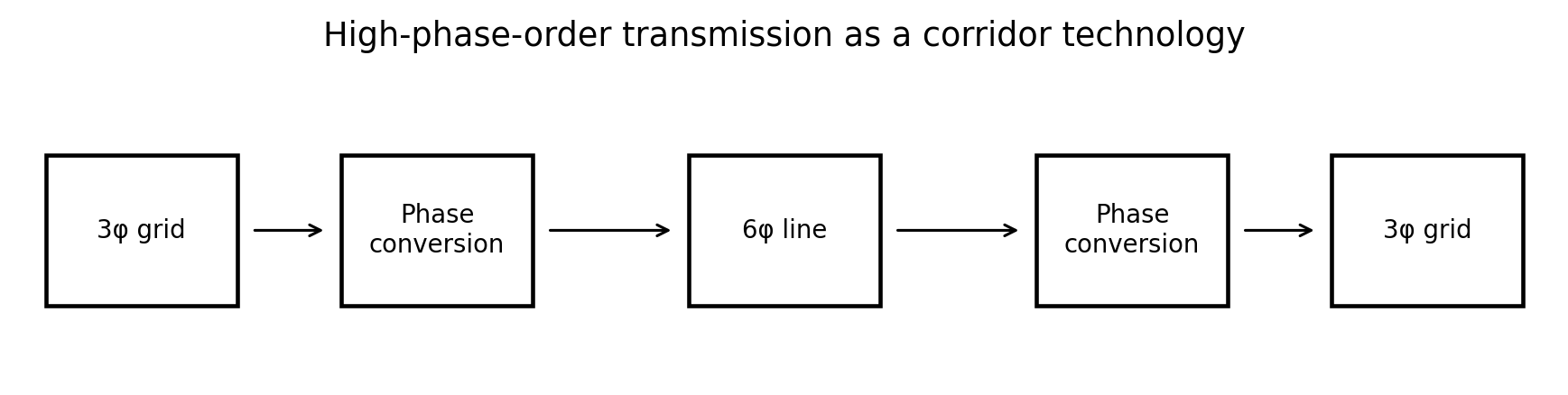}
\caption{Conceptual interface of a six-phase transmission corridor with
a conventional three-phase grid.}\label{fig:hpo}
\end{figure}

A real-world utility-scale HPO commercial demonstration was the Goudey--Oakdale
six-phase pioneering project by the New York State Electric \& Gas Corporation (NYSEG) in 1992 near Binghamton, New York.
Published papers describe reconfiguration of an existing 115-kV
double-circuit three-phase line between Goudey and Oakdale as a
six-phase line and the associated conversion, insulation, and protection
work \autocite{brown1991,stewart1992,rebbapragada1992}. The DOE/OSTI
final report records that the six-phase demonstration line was energized
on July 1, 1992 and operated as an integrated utility tie
\autocite{nyseg1997}.

The six-phase line operated at 93 kV phase-to-ground; because adjacent phases are \(60^\circ\) apart, the adjacent phase-to-phase voltage was also 93 kV, while same phases separated by \(120^\circ\) would have approximately 161 kV phase-to-phase voltage. Thus, the voltage of 93 kV should not be interpreted
as a direct voltage reduction comparable to 115 kV with usual three-phase line-to-line
convention because the voltage definitions and phase relationships
differ. 

The relevant engineering question was whether the existing
six-conductor structure could transfer more useful power within its
corridor and insulation envelope after reconfiguration.

Pre-demonstration studies reported design-specific increases in thermal
capability and surge-impedance loading for the reconfigured line
\autocite{stewart1992}. These values are evidence that a specific HPO
design could improve corridor utilization; they are not universal
multipliers applicable to arbitrary three-phase lines.

Protection was a substantial part of the demonstration. Six phase
creates more possible combinations of faulted and open phases than the
familiar three-phase taxonomy. Rebbapragada \emph{et al.} described a
protection architecture that used available relaying technology together
with logic tailored to the six-phase line \autocite{rebbapragada1992}.
Subsequent reporting on relay operation documented practical experience
with faults on the demonstration line \autocite{apostolov1996}. The
significant conclusion is not that protection became impossible, but
that feasible six-phase protection required additional engineering and
nonstandard system logic.

\begin{longtable}[]{@{}
  >{\raggedright\arraybackslash}p{(\columnwidth - 4\tabcolsep) * \real{0.3333}}
  >{\raggedright\arraybackslash}p{(\columnwidth - 4\tabcolsep) * \real{0.3333}}
  >{\raggedright\arraybackslash}p{(\columnwidth - 4\tabcolsep) * \real{0.3333}}@{}}
\caption{Qualitative summary of the NYSEG HPO demonstration. Exact
ratings and operating details should be taken from the cited project
papers and final report. }\tabularnewline
\toprule\noalign{}
\begin{minipage}[b]{\linewidth}\raggedright
Attribute
\end{minipage} & \begin{minipage}[b]{\linewidth}\raggedright
Existing/host three-phase system
\end{minipage} & \begin{minipage}[b]{\linewidth}\raggedright
Six-phase demonstration concept
\end{minipage} \\
\midrule\noalign{}
\endfirsthead
\toprule\noalign{}
\begin{minipage}[b]{\linewidth}\raggedright
Attribute
\end{minipage} & \begin{minipage}[b]{\linewidth}\raggedright
Existing/host three-phase system
\end{minipage} & \begin{minipage}[b]{\linewidth}\raggedright
Six-phase demonstration concept
\end{minipage} \\
\midrule\noalign{}
\endhead
\bottomrule\noalign{}
\endlastfoot
Corridor conductor count & Six conductors as two 3-phase circuits & Six
conductors as one 6-phase circuit \\
Host-system voltage class & 115-kV class & Approximately 93-kV six-phase
nominal definition \\
Grid interfaces & Conventional 3-phase substations & Phase-conversion
transformer interfaces \\
Principal objective & Conventional double-circuit service & Increased
utilization of existing corridor/structure \\
Protection & Standard 3-phase practices & Modified multi-relay and
six-phase logic \\
Demonstration status & Mature commercial standard & Utility
demonstration, technically successful but not widely replicated \\
\end{longtable}

However, why successful demonstration of the NYSEG Goudey-Oakdale project did not imply widespread adoption? A transmission-uprating technology competes against alternatives, not
merely against an unchanged old line. By the 1990s and afterward,
utilities could consider reconductoring, higher-temperature conductors,
rebuilding at a higher voltage, series compensation, FACTS, additional
circuits, or HVDC in suitable corridors. HPO conversion had to justify
phase-conversion transformers, specialized switching and protection,
maintenance training, spares, and reduced interchangeability with the
rest of the system. Economic studies therefore treated HPO as a specific
planning alternative rather than an inevitable successor to three phase
\autocite{landers1998,hpoecon1984}.

These historical cases establish that several phase orders were technically viable. The next section separates their implementation-specific advantages from properties that follow directly from phase geometry.

\section{III. General Mathematical Properties of Balanced Multiphase Systems}\label{iii.-general-mathematical-properties-of-balanced-multiphase-systems}

A unified formulation helps generalize the calculations of powers and voltages with a balanced multiphase system beyond a three-phase system. 

\subsection{A. Cancellation Condition for Constant Power}
Consider \(m\)-phase
sinusoidal voltages and currents,

\begin{equation}
v_k(t)=\sqrt2V\cos(\omega t-\theta_k), \qquad
i_k(t)=\sqrt2I\cos(\omega t-\theta_k-\phi),
\label{eq:mphase-vi}
\end{equation}

where \(k=0,1,\ldots,m-1\). The instantaneous power in phase
\(k\) is
\begin{equation}
\begin{aligned}
p_k(t)&=v_k(t)i_k(t) \\&=VI\cos\phi+VI\cos(2\omega t-2\theta_k-\phi).
\end{aligned}
\label{eq:mphase-phase-power}
\end{equation}

Summing over all phases yields
\begin{equation}
p_\Sigma(t)=mVI\cos\phi+
VI\sum_{k=0}^{m-1}\cos(2\omega t-2\theta_k-\phi).
\label{eq:mphase-total-power}
\end{equation}

The first term is the total real power \(P_\Sigma\) delivered while the second term of the instantaneous power oscillates at the double frequency $2\omega$ and can be represented as the real part of a complex exponential,

\begin{equation}
p_{2\omega}(t)=VI\operatorname{Re}\left\{e^{j(2\omega t-\phi)}
\sum_{k=0}^{m-1}e^{-j2\theta_k}\right\}.
\label{eq:mphase-power-2omega}
\end{equation}

Therefore, under the sinusoidal assumptions of \eqref{eq:mphase-vi},
aggregate instantaneous power is constant if

\begin{equation}
\sum_{k=0}^{m-1} e^{-j2\theta_k}=0.
\label{eq:power-cancel-condition}
\end{equation}

Equation~\eqref{eq:power-cancel-condition} shows that cancellation of the double-frequency component is a property of the phase-angle set, not a privilege reserved for three phases. For the conventional symmetric \(m\)-phase set defined by 
\begin{equation}
\theta_k=\frac{2\pi k}{m}, \qquad  m>2
\label{eq:symmetric-phase-angles}
\end{equation}
equation~\eqref{eq:power-cancel-condition} becomes a finite geometric series with the double-angle
  phasors sum to zero. This cancellation occurs automatically for every \(m>2\). 
  
  Interestingly, true two-phase power as a special \(m=2\) case uses quadrature angles \(0\) and \(\pi/2\), instead, in \eqref{eq:two-phase-alpha} and \eqref{eq:two-phase-beta},  which is an alternative angular construction \(\theta_k=\frac{\pi k}{m}
\) and satisfies the power-cancellation condition for every \(m>1\), because the corresponding double-angle phasors form the \(m\) roots of unity: \(
\sum_{k=0}^{m-1} e^{-j2\theta_k}=\sum_{k=0}^{m-1} e^{-j2\pi k/m}=0.\) However, this alternative angular construction does not result in conventional balanced \(m\)-phase voltage sets for $m>2$. Obviously, single phase has only one term and cannot satisfy the cancellation
  condition by any angular construction.

This derivation also indicates the assumptions behind the familiar
result. Harmonics, unequal phase amplitudes, unequal load impedances,
phase-angle errors, and unbalanced faults can reintroduce oscillatory
components in aggregate power. Thus ``constant power'' is an ideal, balanced
sinusoidal property, not an assertion that real three-phase grids have
no power oscillations.

\subsection{B. Adjacent-Phase Voltage Differences}

For a symmetric \(m\)-phase set, write the voltage phasor of each phase as:

\begin{equation}
\mathbf V_k=V_{ph}e^{j2\pi k/m}.
\label{eq:mphase-voltage-phasor}
\end{equation}

The voltage difference between phases separated by \(r\) phase steps is

\begin{equation}
\begin{aligned}
|\mathbf V_k-\mathbf V_{k+r}|
&=V_{ph}|1-e^{j2\pi r/m}|\\
&=2V_{ph}\left|\sin\left(\frac{\pi r}{m}\right)\right|.
\end{aligned}
\label{eq:phase-difference-r}
\end{equation}

For adjacent phases, \(r=1\),

\begin{equation}
V_{adj}=2V_{ph}\sin\left(\frac{\pi}{m}\right).
\label{eq:adjacent-phase-voltage}
\end{equation}

Equation~\eqref{eq:adjacent-phase-voltage} is central to HPO line compaction. If physically adjacent
conductors are ordered by electrical phase angle, increasing \(m\)
reduces the voltage difference between adjacent conductors relative to
phase-to-neutral voltage. This may allow closer spacing for a given
local field criterion, although actual insulation design also depends on
overvoltages, conductor-to-ground clearances, switching surges,
lightning performance, corona, and environmental constraints
\autocite{stewart1978b,stewart1992}.

\begin{longtable}[]{@{}rrr@{}}
\caption{Adjacent-phase voltage ratio in a symmetric phase set.}
\tabularnewline
\toprule\noalign{}
Phase order \(m\) & Adjacent phase angle & \(V_{adj}/V_{ph}\) \\
\midrule\noalign{}
\endfirsthead
\toprule\noalign{}
Phase order \(m\) & Adjacent phase angle & \(V_{adj}/V_{ph}\) \\
\midrule\noalign{}
\endhead
\bottomrule\noalign{}
\endlastfoot
3 & \(120^\circ\) & \(\sqrt{3}=1.732\) \\
6 & \(60^\circ\) & \(1.000\) \\
9 & \(40^\circ\) & \(2\sin20^\circ=0.684\) \\
12 & \(30^\circ\) & \(2\sin15^\circ=0.518\) \\
\end{longtable}

\begin{figure}
\centering
\includegraphics[width=0.76\textwidth,height=\textheight]{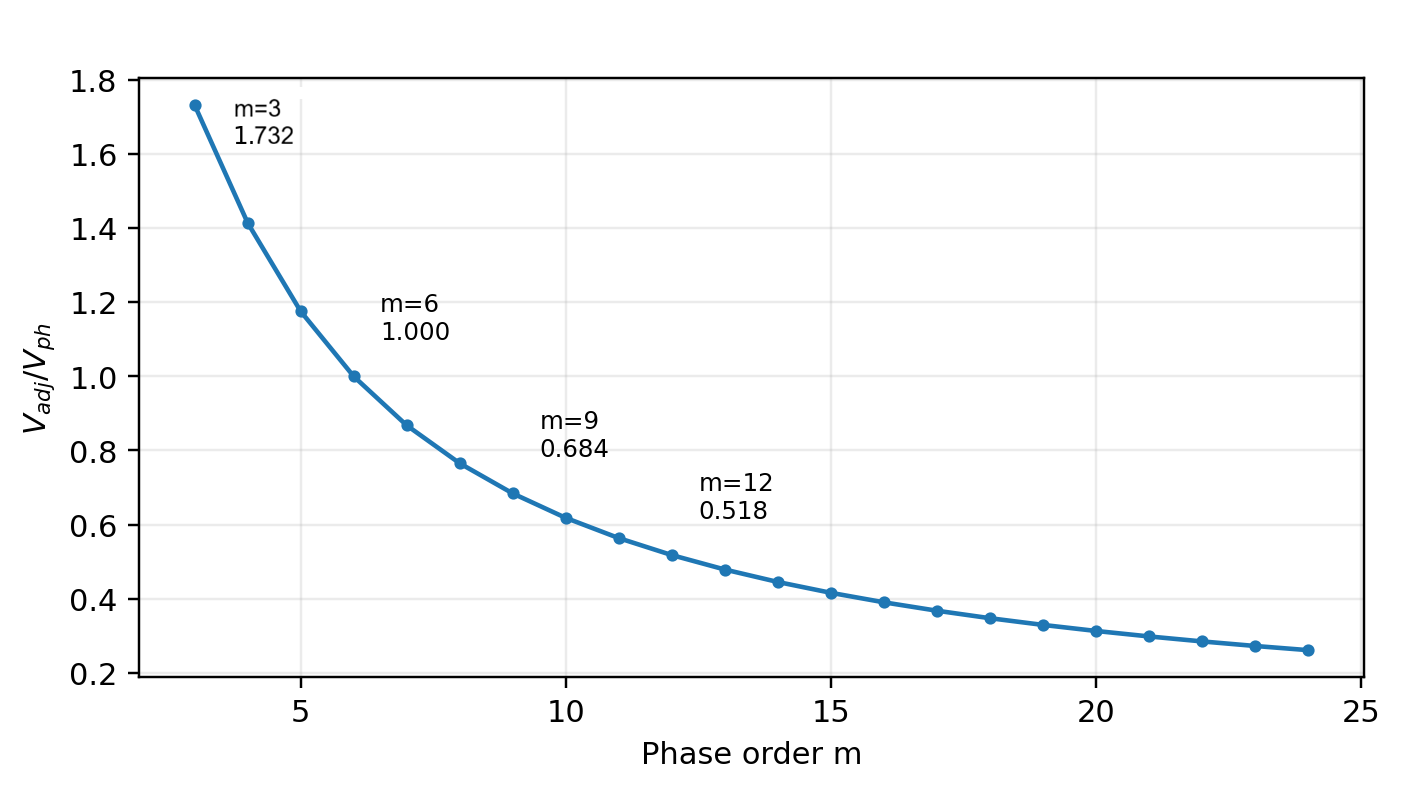}
\caption{Adjacent-phase voltage ratio versus phase order.}\label{fig:vadj}
\end{figure}

The table also reveals diminishing geometric return. Increasing \(m\)
continues to reduce adjacent-phase voltage, but each added phase also
adds a complete conductor, terminal, measurement, and switching burden.
For large \(m\),

\begin{equation}
2\sin(\pi/m)\approx\frac{2\pi}{m},
\label{eq:adjacent-phase-asymptotic}
\end{equation}

so the local geometric benefit scales approximately as \(1/m\), while
many equipment counts scale at least linearly with \(m\).

\subsection{C. Would High-Phase Order Make Power-System Mathematics
Simpler?}\label{viii.-would-six-phases-make-power-system-mathematics-easier}

Let us use six-phase transmission as an example. For six equally spaced phases, adjacent phases differ by \(60^\circ\),
so from~\eqref{eq:phase-difference-r},

\begin{equation}
V_{adj}=2V_{ph}\sin30^\circ=V_{ph}.
\label{eq:six-phase-adjacent-voltage}
\end{equation}

Thus one selected voltage ratio is indeed simpler than the three-phase
relation \(V_{LL}=\sqrt{3}V_{ph}\). But this does not simplify the
physical system overall for four reasons. First, \(\sqrt{3}\) is variable-dependent. Three-phase power written in phase quantities is \(P_\Sigma=3V_{ph}I_{ph}\cos\phi\), with no square root.
The factor appears when line-to-line voltage is chosen because of phasor
geometry. Per-unit normalization and \(dq\) transformations routinely
absorb such constants into base quantities or transformation matrices. Second, a six-phase set does not eliminate irrational voltage ratios. It
contains phase pairs separated by \(60^\circ\), \(120^\circ\), and
\(180^\circ\), giving voltage differences of \(V_{ph}\),
\(\sqrt{3}V_{ph}\), and \(2V_{ph}\) respectively. The constant has
disappeared from one adjacency relation, not from the system. Third, phase-space dimension increases. Fortescue's
symmetrical-component framework decomposes polyphase quantities into
modal components \autocite{fortescue1918}. A general \(m\)-phase vector
can be transformed through the discrete-Fourier structure

\begin{equation}
X_q=\frac{1}{m}\sum_{k=0}^{m-1}x_k e^{-j2\pi qk/m},
\quad q=0,\ldots,m-1.
\label{eq:dft-modal-components}
\end{equation}

Three phase therefore has three phase variables and the familiar zero-,
positive-, and negative-sequence structure. A six-phase system has six
phase variables and correspondingly more independent modal degrees of
freedom. Modern multiphase-machine theory often reorganizes these into a
torque-producing fundamental subspace plus additional orthogonal
subspaces and zero-sequence components \autocite{levi2008,rockhill2015}.
Those extra subspaces can be useful, especially for fault-tolerant
drives, but they are additional degrees of freedom rather than a
mathematical simplification.

Fourth, hardware and control dimensions track physical channels. A
six-phase controller may require more current measurements, modulation
states, fault classifications, and switching constraints. Modern
processors incur negligible cost multiplying by \(\sqrt{3}\); physical
channels, sensors, breakers, and protection states are not negligible.

The appropriate conclusion for a conventional electromechanical grid is
therefore precise: \textbf{six phases simplify selected local phasor
ratios but do not automatically simplify the overall analysis, control,
protection, or equipment design of a transmission system.} This
conclusion, however, treats additional physical channels primarily as
costs. If future grid interfaces are converter dominated, the same added
dimensions can become control resources. That alternative assumption is
examined in Section IV.

\subsection{D. Why not 9, 12, or more phases?}\label{x.-why-not-nine-twelve-or-more-phases}

Higher phase orders continue the HPO geometric trend, but the marginal
value decreases while the system departs further from the standardized
three-phase ecosystem. Six phase is especially interesting because an
existing double-circuit three-phase structure already has six
conductors. Nine phase typically requires three additional phase
conductors relative to six; twelve phase can map naturally to four
three-phase groups but requires an even larger conversion and switching
structure.

A simple combinatorial illustration is the number of unordered phase
pairs,

\begin{equation}
N_{pair}=\binom{m}{2}=\frac{m(m-1)}{2}.
\label{eq:phase-pair-count}
\end{equation}

This quantity is \textbf{not} a model of protection cost, because real
relay complexity depends on grounding, fault categories, line topology,
breaker arrangement, and chosen algorithms. It is nevertheless a useful
reminder that the number of pairwise phase relationships grows faster
than the phase count itself:

\begin{equation}
m=3\Rightarrow N_{pair}=3,
\quad m=6\Rightarrow15,
\quad m=9\Rightarrow36,
\quad m=12\Rightarrow66.
\label{eq:phase-pair-examples}
\end{equation}

The broader equipment burden grows through conductors, bushings,
terminals, instrument transformers or sensors, breaker poles,
disconnects, surge arresters, buswork, control I/O, test procedures, and
spares. Meanwhile, the adjacent-voltage benefit of \eqref{eq:adjacent-phase-voltage} becomes
progressively smaller in absolute terms as \(m\) rises. Economic studies
of HPO therefore focused strongly on six phase and, in some comparisons,
twelve phase rather than suggesting that arbitrarily large phase orders
would be optimal \autocite{hpoecon1984}.

The preceding analysis treats additional phases largely as additional system dimensions. The next Section IV analyzes how the interpretation changes when those dimensions become electronically controllable.

\begin{figure}
\centering
\includegraphics[width=0.78\textwidth,height=\textheight]{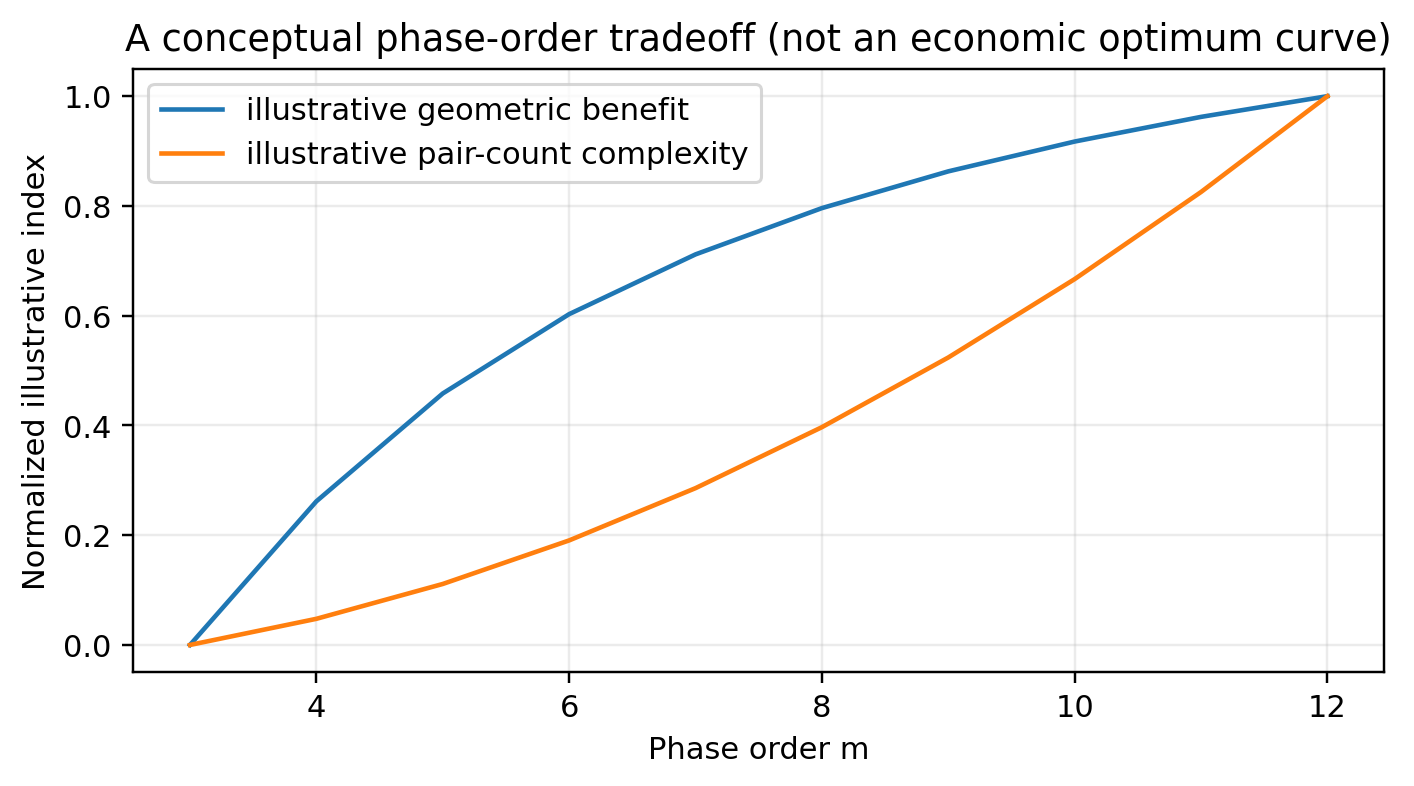}
\caption{Conceptual illustration of decreasing geometric benefit and
increasing combinatorial complexity with phase order. The curves are
normalized illustrative indices, not an economic optimum
calculation.}\label{fig:tradeoff}
\end{figure}

\section{IV. High Phase Orders in a
Power-Electronics-Dominated Grid}\label{ix.-when-higher-phase-order-becomes-an-asset-a-power-electronics-dominated-grid}

The historical three-phase optimum was selected in a grid built around
synchronous machines, magnetic transformers, mechanically operated
switchgear, and protection schemes whose hardware count grew directly
with phase count. A future grid dominated by modular power-electronic
interfaces changes that cost function. Converter legs can be replicated,
phase displacement and phase conversion can increasingly be synthesized electronically, reducing reliance on specialized magnetic phase-conversion structures,
current measurements are inexpensive, and fault accommodation can be
embedded in control. Under that counterfactual assumption, \(m>3\) has
several theoretically serious advantages that should be separated from
the superficial claim that six phase is attractive merely because one
voltage ratio loses \(\sqrt{3}\).

\subsection{A. Advantage 1: modular decomposition into interleaved
three-phase
subsystems}\label{a.-advantage-1-modular-decomposition-into-interleaved-three-phase-subsystems}

A symmetrical six-phase set has

\begin{equation}
\theta_k=\frac{k\pi}{3},\qquad k=0,\ldots,5.
\label{eq:six-phase-angle-set}
\end{equation}

The even-indexed phases form \(\{0^\circ,120^\circ,240^\circ\},\) and the odd-indexed phases form \(\{60^\circ,180^\circ,300^\circ\}.\) Thus a six-phase system can be regarded as two balanced three-phase
groups displaced by \(60^\circ\). More generally, any symmetric phase
order \(m=3q\) can be partitioned into \(q\) interleaved balanced
three-phase sets. This factorization is not merely a visual phasor
identity. Portela and Tavares showed that an appropriate component
transformation for a six-phase transmission line can represent the line
by two uncoupled three-phase elements under the symmetry conditions of
their formulation \autocite{portela1993}. In a converter-dominated
implementation, the same structure suggests modular realization: a
six-phase terminal can be assembled from two familiar three-leg
converter groups, each retaining conventional \(dq\) control internally
while a higher-level controller coordinates power sharing.

Six phase also has a second natural partition into three antipodal
pairs, \(\{0^\circ,180^\circ\}\), \(\{60^\circ,240^\circ\}\), and
\(\{120^\circ,300^\circ\}\). The arithmetic identity \(6=2\times3\)
therefore corresponds to useful electrical subgroup structure. It does
not prove that six is globally optimal, but it explains why six phase
can be more modular than an arbitrary prime phase order.

\subsection{B. Advantage 2: additional phases create redundant control
degrees of
freedom}\label{b.-advantage-2-additional-phases-create-redundant-control-degrees-of-freedom}

For a three-wire balanced three-phase system, the constraint \(i_a+i_b+i_c=0\) leaves two independent current coordinates, which are exactly the two
coordinates needed to synthesize the fundamental \(\alpha\beta\)
power-transfer vector. Under the idealized assumptions of a single zero-sum current constraint and a two-dimensional fundamental \(\alpha\beta\) power-transfer subspace, the number of remaining independent current-control degrees of freedom is
\begin{equation}
d_{\mathrm{extra}}=(m-1)-2=m-3.
\label{eq:extra-control-dof}
\end{equation}
This is an idealized dimensional count rather than a universal result for every multiphase topology; additional neutral, winding, converter, or zero-sequence constraints may reduce the available degrees of freedom.

For \(m=6\), this gives three additional coordinates. The exact
decomposition depends on neutral connections, winding layout, and
converter topology, but the key point is robust: higher phase order can
make the system \emph{overactuated}. Multiphase-drive research
explicitly exploits these additional degrees of freedom for fault
tolerance, current sharing, harmonic control, and other objectives
\autocite{levi2016,levi2008}.

This allows post-contingency operation to be formulated as constrained
optimization rather than as a binary healthy/failed state. For example,
after an open-phase event one can seek

\begin{equation}
\begin{aligned}
&\min_{\mathbf i}\; \mathbf i^{T}R\mathbf i \\
\text{subject to} \qquad &P=P^{\star},\qquad Q=Q^{\star},\qquad i_f=0,\qquad |i_k|\le I_{\max}.
\end{aligned}
\label{eq:postfault-constraints}
\end{equation}

A three-phase converter has little remaining freedom after losing one
phase. A six- or nine-phase interface can redistribute current among
healthy phases while preserving the required fundamental power vector.
Under modern converter control, the extra dimensions are therefore not
only modeling burden; they are control authority.

\subsection{C. Advantage 3: higher dimensionality can coexist with
simpler modal
structure}\label{c.-advantage-3-higher-dimensionality-can-coexist-with-simpler-modal-structure}

The usual objection that six phase has six variables while three phase
has only three is correct in phase coordinates but incomplete in modal
coordinates. For an ideally symmetric line, the phase impedance or
admittance matrix approaches circulant or block-circulant structure. The
discrete Fourier matrix \(\mathbf F_m\) diagonalizes a circulant matrix,

\begin{equation}
\mathbf Z_{\mathrm{modal}}=\mathbf F_m\mathbf Z_{\mathrm{phase}}\mathbf F_m^{-1},
\label{eq:modal-impedance-transform}
\end{equation}

so the physical phase variables separate into independent propagation
modes. Portela and Tavares' six-phase formulation is a particularly
strong example because the transformed line can be represented as two
uncoupled three-phase elements \autocite{portela1993}. Classical
six-phase fault-analysis papers likewise developed symmetrical-component
and modal formulations rather than solving every fault directly in six
coupled phase variables \autocite{bhatt1977,badawy1991}.

This does \textbf{not} mean every six-phase fault is easier than the
corresponding three-phase fault. An asymmetrical fault breaks phase
symmetry at its boundary conditions and can couple transformed
components. The defensible theoretical statement is narrower:
\textbf{increased phase count increases dimensionality, but increased
phase symmetry can simultaneously increase block diagonalization and
reuse of standard three-phase solution structures.} Therefore
computational difficulty need not scale directly with \(m\).

\subsection{D. Advantage 4: greater transferable power per constrained
corridor}\label{d.-advantage-4-greater-transferable-power-per-constrained-corridor}

The most fundamental line-level advantage remains geometry. For a
symmetric \(m\)-phase set, the voltage difference between adjacent
ordered phases is

\begin{equation}
V_{\mathrm{adj}}=2V_{ph}\sin\left(\frac{\pi}{m}\right).
\label{eq:adjacent-phase-voltage-corridor}
\end{equation}

Suppose, as an intentionally idealized limiting case, that
adjacent-conductor voltage stress is the binding corridor constraint and
that the allowed adjacent difference is \(V_{\mathrm{adj,max}}\). Then

\begin{equation}
V_{ph,\max}=\frac{V_{\mathrm{adj,max}}}{2\sin(\pi/m)}.
\label{eq:max-phase-voltage-field-limit}
\end{equation}

If each of the \(m\) conductors can carry RMS current \(I\) at unity
power factor, the corresponding apparent-power scaling is

\begin{equation}
S_{\max}\propto\frac{m}{2\sin(\pi/m)}.
\label{eq:idealized-capacity-scaling}
\end{equation}

For large \(m\), \(\sin(\pi/m)\approx\pi/m\), so this idealized metric
grows approximately as \(m^2/(2\pi)\). The result must not be
interpreted as a practical \(m^2\) law for transmission capacity:
conductor-to-ground insulation, nonadjacent phase voltages, corona,
thermal current limits, mechanical spacing, stability, and tower
geometry intervene. Its value is conceptual. It isolates the electrical-stress mechanism by which HPO can contribute to increased corridor power density. Actual MW per unit right-of-way additionally depends on conductor count, mechanical spacing, phase-to-ground insulation, corona, tower geometry, and thermal limits. The NYSEG
demonstration and earlier feasibility studies were motivated by this
corridor-utilization mechanism \autocite{stewart1978a,brown1991}.

\subsection{E. Advantage 5: compact HPO geometry can improve natural
loading and
loadability}\label{e.-advantage-5-compact-hpo-geometry-can-improve-natural-loading-and-loadability}

Phase order also changes the distributed electromagnetic parameters of
the line. For a dominant propagation mode, the surge impedance is
approximately determined by the modal inductance and capacitance:
\begin{equation}
Z_c\approx\sqrt{\frac{L}{C}},
\label{eq:surge-impedance}
\end{equation}

and the surge-impedance loading scales approximately as

\begin{equation}
P_{\mathrm{SIL}}\propto\frac{V^2}{Z_c}.
\label{eq:sil-scaling}
\end{equation}
For a fixed voltage convention, natural loading increases as the relevant modal characteristic impedance decreases.

Compact HPO geometry can alter modal \(L\) and \(C\) in a direction that
reduces characteristic impedance and increases natural loading and
transfer capability. HPO design studies explicitly treated minimum
impedance, improved surge loading, and improved power transfer as design
objectives, and separate work investigated the stability and loadability
of six-phase/HPO systems
\autocite{chandrasekaran1986,tiwari1995}. These are
design-dependent effects rather than universal ratios, but they provide
a second capacity mechanism beyond simple thermal ampacity. For corridor
planning, a relevant theoretical metric is therefore not only conductor
efficiency but

\begin{equation}
\frac{\text{stable transferable MW}}{\text{right-of-way width}}.
\label{eq:stable-mw-per-row}
\end{equation}

\subsection{F. Advantage 6: more phase symmetry provides additional
cancellation
opportunities}\label{f.-advantage-6-more-phase-symmetry-provides-additional-cancellation-opportunities}

Uniform phase sets obey the roots-of-unity identity

\begin{equation}
\sum_{k=0}^{m-1}
e^{j2\pi r k/m}
=
\begin{cases}
m, & r\equiv 0 \pmod m,\\
0, & r\not\equiv 0 \pmod m,
\end{cases}
\qquad r\in\mathbb{Z}.
\end{equation}

This identity underlies cancellation of selected space and time
harmonics in balanced multiphase systems. When converters can
independently shape phase waveforms, additional phase groups provide
more freedom to interleave switching, distribute ripple, suppress
selected harmonic components, and reduce per-phase current stress
\autocite{levi2016}. Similar spatial cancellation can be exploited in
overhead-line geometry. General transmission-line design must account for electric and magnetic
field levels as corridor constraints \autocite{stoffel1994}. HPO research
demonstrated that appropriate high-phase-order conductor arrangements can
reduce ground-level magnetic fields while transmitting power in a compact
corridor \autocite{stewart1993field}. Thus additional phase symmetry can improve
not only electrical power transfer but also the \textbf{electromagnetic
footprint per transmitted MW}.

These six advantages are summarized in the following table.

\begin{longtable}[]{@{}
  >{\raggedright\arraybackslash}p{(\columnwidth - 6\tabcolsep) * \real{0.2500}}
  >{\raggedright\arraybackslash}p{(\columnwidth - 6\tabcolsep) * \real{0.2500}}
  >{\raggedright\arraybackslash}p{(\columnwidth - 6\tabcolsep) * \real{0.2500}}
  >{\raggedright\arraybackslash}p{(\columnwidth - 6\tabcolsep) * \real{0.2500}}@{}}
\caption{Six theoretically serious advantages of phase order above three
when power-electronic conversion and protection are assumed to remove
much of the historical hardware penalty.}\tabularnewline
\toprule\noalign{}
\begin{minipage}[b]{\linewidth}\raggedright
Theoretical advantage of \(m>3\)
\end{minipage} & \begin{minipage}[b]{\linewidth}\raggedright
Underlying structure
\end{minipage} & \begin{minipage}[b]{\linewidth}\raggedright
Potential value in a converter-dominated grid
\end{minipage} & \begin{minipage}[b]{\linewidth}\raggedright
Main qualification
\end{minipage} \\
\midrule\noalign{}
\endfirsthead
\toprule\noalign{}
\begin{minipage}[b]{\linewidth}\raggedright
Theoretical advantage of \(m>3\)
\end{minipage} & \begin{minipage}[b]{\linewidth}\raggedright
Underlying structure
\end{minipage} & \begin{minipage}[b]{\linewidth}\raggedright
Potential value in a converter-dominated grid
\end{minipage} & \begin{minipage}[b]{\linewidth}\raggedright
Main qualification
\end{minipage} \\
\midrule\noalign{}
\endhead
\bottomrule\noalign{}
\endlastfoot
Modular decomposition & \(6\phi=2\times3\phi\); more generally \(m=3q\)
& Reuse of standard three-leg converter and \(dq\) modules & Exact
decoupling depends on topology and line symmetry \\
Redundant control DOF & Fundamental power uses a 2-D subspace &
Fault-tolerant current redistribution, thermal balancing, secondary
objectives & Available DOF depend on neutral/winding constraints \\
Structured modal analysis & DFT diagonalizes circulant/block-circulant
matrices & Block-diagonal models and reuse of familiar 3-phase methods &
Fault boundary conditions can recouple modes \\
Corridor power density & \(V_{adj}=2V_{ph}\sin(\pi/m)\) & Higher
voltage/power utilization within a constrained geometry & Phase-ground
insulation, corona, mechanics, and thermal limits remain \\
Natural loading/loadability & HPO geometry modifies modal \(L\), \(C\),
and \(Z_c\) & Higher SIL and potentially greater stable MW/ROW &
Strongly design-specific \\
Harmonic/field cancellation & Roots-of-unity symmetry and spatial phase
ordering & Ripple shaping, harmonic cancellation, lower external
magnetic field & Requires coordinated geometry/control \\
\end{longtable}

\subsubsection{A necessary counter-result: more phases do not
intrinsically reduce conductor
loss}\label{a-necessary-counter-result-more-phases-do-not-intrinsically-reduce-conductor-loss}

One tempting claim should be rejected. Assume total transmitted real
power \(P\), phase RMS voltage \(V_{ph}\), power factor \(\cos\phi\),
line length \(\ell\), resistivity \(\rho\), and total conductor
cross-sectional area \(A_\Sigma\) are fixed. Dividing the material
equally among \(m\) phase conductors gives \(A_k=A_\Sigma/m\) and

\begin{equation}
R_k=\rho\ell\frac{m}{A_\Sigma}.
\label{eq:phase-conductor-resistance}
\end{equation}

From \(P=mV_{ph}I\cos\phi\),

\begin{equation}
I=\frac{P}{mV_{ph}\cos\phi}.
\label{eq:phase-current-fixed-copper}
\end{equation}

The total conductor loss is then

\begin{equation}
\begin{aligned}
P_{\mathrm{loss}}
=mI^2R_k \quad=\frac{P^2\rho\ell}{A_\Sigma V_{ph}^2\cos^2\phi},
\end{aligned}
\label{eq:fixed-copper-loss}
\end{equation}

Remarkably, the phase order \(m\) does not appear explicitly in this expression. Thus, under the idealized assumptions of equal phase RMS voltage, fixed total conductor cross-sectional area, uniform current sharing, equal conductor length and resistivity, and purely resistive conductor losses, \textbf{increasing the phase order alone does not intrinsically reduce \(I^2R\) loss}. This result neglects AC resistance effects such as skin and proximity effects, as well as corona, dielectric loss, conductor geometry, reactive parameters, and other phase-order-dependent transmission-line effects. Any practical efficiency advantage of higher phase order must therefore arise indirectly through changes in usable operating voltage, conductor geometry, field distribution, current allocation, or other system-design variables.

Thus \textbf{increasing phase count
alone does not improve \(I^2R\) efficiency} under these fixed
conditions. Any efficiency benefit of HPO must arise indirectly--for
example, by allowing a higher usable voltage, a different total
conductor allocation, reduced losses from optimized current sharing, or
different converter/filter requirements. This counter-result is useful
because it separates genuine phase-order advantages from claims that are
merely consequences of changing another design constraint.

The broader theoretical implication is that the optimum phase count is
technology dependent. In the historical grid, extra phases meant extra
electromechanical hardware and nonstandard interfaces. In a
converter-dominated grid, some of those costs shrink while modularity,
redundant actuation, and waveform shaping become more valuable. Three
phase may remain optimal, but that conclusion can no longer be justified
solely by counting relays or breaker poles; it must be demonstrated from
a modern multiobjective optimization.

\subsection{G. Modern Multiphase
Applications}\label{xi.-modern-multiphase-applications}

The dominance of three-phase transmission should not be generalized into
a rule that three phases are best for every electromechanical system.
Multiphase machines with five, six, or more phases can reduce per-phase
current, increase fault tolerance, distribute converter stress, and
maintain torque production after loss of one or more phases
\autocite{levi2008}. These attributes are valuable in ship propulsion,
aerospace systems, traction, and other applications where converter and
machine are designed as one integrated system.

High-pulse-number converters similarly use phase-shifted three-phase
groups to reduce characteristic harmonics. A twelve-pulse rectifier, for
example, exploits two displaced three-phase supplies locally. Such
architectures demonstrate a central distinction: the phase order that is
optimal inside a converter-machine subsystem need not be the phase order
that minimizes cost and complexity across a continental transmission
network.

This distinction also explains why six-phase HPO remains technically
interesting. A constrained right-of-way (ROW) is a localized optimization
problem. If the corridor already contains six conductors and new land is
difficult to acquire, phase reconfiguration may deserve study. But the
boundary conditions at both ends are still a predominantly three-phase
grid, so conversion equipment becomes part of the economic comparison.

\section{\texorpdfstring{V. Discussion: Why \(m=3\) Is a Technological
Sweet
Spot}{V. Discussion: Why m=3 Is a Technological Sweet Spot}}\label{xii.-discussion-why-m3-is-a-technological-sweet-spot}

A useful central proposition is:

\begin{quote}
\textbf{Three phases constitute the lowest practical phase order that
simultaneously provides constant balanced aggregate power, a naturally
rotating field, and an economical symmetric three-conductor architecture
for bulk AC transmission.}
\end{quote}

Two phase satisfies the first
two properties. It therefore refutes any explanation based solely on
constant power or rotating-field production. What it does not offer in
its conventional four-wire realization is the same symmetric
three-conductor bulk-power path. A three-wire two-phase arrangement can
reduce conductor count, but the shared conductor carries nonzero
balanced current and does not recover the cyclic equality of the
three-phase line.

Three phase is also not a mathematical global optimum proven from one
scalar objective. Transmission engineering is multiobjective. If
right-of-way width dominates all other costs, six-phase HPO may
outperform a conventional three-phase reconstruction in a specific
corridor. If fault tolerance, converter current sharing, modal
modularity, or waveform shaping dominate, a six- or higher-phase
interface may be superior. If the load is small and mostly single phase,
a two-wire supply is preferable. The defensible historical claim is
therefore that for a general interconnected \textbf{electromechanical}
AC bulk-power network, three-phase transmission sits at an unusually favorable point
where major polyphase benefits have already been obtained but
phase-order hardware complexity remains minimal. Section IV shows why
this optimum is less obvious for a future converter-dominated network.

This conclusion has two historical layers.

\textbf{Physical and economic selection.} Three-phase transmission offered constant
balanced aggregate power, a natural rotating field, strong conductor
utilization, and convenient machine and transformer structures. The
Niagara experience is especially instructive because it placed two-phase
generation and three-phase transmission in the same project
\autocite{ethwBuffalo}.

\textbf{Path dependence and network effects.} Once three-phase
generation, transformation, switching, relaying, metering, and
utilization became standardized, the installed base magnified the
original advantage. New equipment became cheaper because it was produced
at scale; protection philosophy and training assumed three phases;
interconnection compatibility rewarded conformity. By the time HPO
transmission was technically mature enough for utility demonstration, it
had to overcome not only engineering tradeoffs but also a century of
standardization.

The result can be summarized as a phase-order progression. One phase is
hardware-minimal but has double-frequency aggregate-power pulsation. Two
phase removes that pulsation and produces an excellent rotating field.
Three phase retains those benefits in an elegant three-conductor
symmetric architecture. Higher phase orders do not remove a fundamental
defect of balanced three-phase power, but they can add new
resources--corridor compaction, modular three-phase subgroup structure,
redundant control coordinates, modal separability, greater natural
loading, and harmonic/field cancellation. Whether those resources
outweigh the remaining costs is increasingly a technology-dependent
question rather than a fixed consequence of nineteenth-century hardware.

\section{VI. Conclusions}\label{xiii.-conclusions}

This paper revisited the familiar three-phase grid as the outcome of a
design competition rather than as an assumed starting point. Several
conclusions follow.

First, single-phase AC is fully capable of transmitting useful power, as
early installations demonstrated, but balanced single-phase
instantaneous power contains a double-frequency component.

Second, true balanced two-phase power does \textbf{not} share that
aggregate-power pulsation. Its two quadrature phase powers cancel the
\(2\omega\) component exactly, and two-phase windings can generate a
constant rotating field. Therefore the common claim that three phase
displaced two phase because only three phase provides constant power is
incorrect.

Third, the stronger reason for three-phase dominance is system
architecture. Three phase combines constant balanced power and
rotating-field operation with a symmetric three-wire bulk-power path in
which the balanced phase currents sum to zero. Conductor-economy
percentages depend on the comparison constraints, but the
three-conductor cyclic symmetry itself is robust.

Fourth, higher phase order is technically meaningful. The general
relation \(V_{adj}=2V_{ph}\sin(\pi/m)\) explains why HPO transmission
can compact conductor geometry and improve right-of-way utilization. The
NYSEG Goudey--Oakdale project demonstrated that six-phase transmission
could be converted, protected, switched, and operated within a
conventional utility system \autocite{nyseg1997}. Its success
established feasibility, not inevitability: conversion equipment,
protection complexity, nonstandard hardware, and competition from other
uprating technologies limited replication.

Fifth, removing \(\sqrt{3}\) from one six-phase adjacent-voltage
relation does not by itself simplify the grid. The square root is only a
geometric scaling factor associated with the chosen three-phase voltage
variables. Higher phase order introduces additional physical variables
and modal subspaces, while modern computation makes constant factors
essentially irrelevant.

Sixth, those additional dimensions should not be classified only as
complexity. Under a power-electronics-dominated architecture, selected higher phase orders—especially \(m=3q\)—can provide modular decomposition into interleaved three-phase groups, redundant control degrees of freedom, structured
modal models, increased corridor power density, potentially higher
natural loading, and additional harmonic/field cancellation. Conversely,
a fixed-copper derivation shows no intrinsic \(I^2R\) efficiency gain
from phase count alone. These results make the phase-order optimum
conditional on technology and design constraints.

The broad conclusion is therefore not that three is the only viable
phase order or that it is universally optimal. Rather, \textbf{for the conventional electromechanical AC grid, three phases captured the principal electromagnetic advantages of polyphase power at nearly the minimum practical system complexity for an interconnected bulk-power network.} Physics and economics placed three
phase on the shortlist; industrial standardization and network effects
made that choice extraordinarily persistent.

\printbibliography[title=References]

\end{document}